**Assessing flood-conditioned power dependency of urban rail transit and its effects on resilience under climate change**

Wei Bi [1,*], Bryan T. Adey [1], Jürgen Hackl [2]
[1] Institute of Construction and Infrastructure Management, Department of Civil, Environmental and Geomatic Engineering, ETH Zurich, Zurich 8093, Switzerland
[2] Department of Civil and Environmental Engineering, Princeton University, Princeton, NJ, 08544, United States

[*] Corresponding author. Email address: bi@ibi.baug.ethz.ch (W. Bi).

**Abstract:** Urban rail transit (URT) faces increasing flood risk while relying on traction power systems whose failures can trigger cascading disruption. Existing studies on URT flood resilience often oversimplify power dependency by overlooking traction power redundancy and rarely consider flood-conditioned cascading impacts, providing limited insight into how flood-induced traction power failures propagate through operations and affect system resilience. This study marks the first quantitative assessment of flood-conditioned power dependency of URT and its effects on system resilience. The methodology integrates a double-layer URT network model, spatial flood exposure assessment, service disruption simulation, a flood-conditioned power dependency index, and a journey-based performance metric that captures resilience properties of robustness and redundancy. It is applied to the London rail transit network under surface water flood scenarios across current and RCP8.5 climate conditions, two traction power feeding mechanisms, and four substation flood-failure thresholds. Results show that traction power redundancy fundamentally shapes the severity of cascading impacts. Under double-end feeding, power-attributable performance loss remains minimal, whereas single-end feeding leads to substantially greater loss and is highly sensitive to substation failure thresholds. The proposed dependency index is strongly correlated with dependency-related performance loss, supporting its use as a proxy for dependency-induced additional consequences in strategic planning.

## 1 Introduction

Modern transport systems face challenges that are different from those encountered by the pioneering engineers who built them in earlier centuries. Two pressing challenges are the increasing water-related hazard risks posed by climate change [1] and the growing system dependencies among various domains driven by electrification and digitalisation [2]. Multimodal urban rail transit (URT) systems, such as metros, light rail, and trams, are energy-efficient socio-technical systems that play a vital role in sustaining urban mobility [3], yet they exemplify such complexities and are vulnerable to floods. In recent years, URT systems in many cities worldwide have experienced severe rainfall-induced flash flood incidents, where rainfall totals comparable to a typical monthly amount can fall within only a few hours [4]. Examples include New York in 2012, 2021, 2023, and 2025; Brussels in 2016; Sydney in 2020 and 2025; London and Zhengzhou in 2021; Milan, Madrid, and Hong Kong in 2023; Valencia in 2024; Paris in 2025; and Zurich in 2026 [5–9]. These events have caused significant equipment damage, extensive service suspensions, and, in some cases, passenger fatalities when trains were trapped in flooded tunnels. Beyond directly disrupting URT operations by inundating rail assets such as stations and tracks, rainfall-induced flooding can also trigger indirect disruptions by damaging power substations and interrupting electricity supply [10,11]. Such functional dependency-driven cascading failures create additional vulnerability to URT systems and are receiving growing attention in transport resilience and adaptation planning [12,13], particularly as future climate conditions introduce greater uncertainty in the frequency, intensity, and spatial extent of extreme rainfall events.

Incorporating power dependency into URT flood resilience assessments is therefore important for capturing the effects of cross-sectoral, network-wide flood risks. In the context of infrastructure systems, resilience refers to the ability to withstand adverse conditions, maintain service during disruption, and recover rapidly following disruption [14,15]. Assessing infrastructure resilience involves evaluating system performance from disruption onset to restoration completion, with more resilient systems experiencing less performance loss over this period [16]. Some studies, however, adopt a robustness-oriented perspective and assess only the immediate performance degradation following disruption, without explicitly considering subsequent recovery. While this approach does not represent the full disruption-recovery process, it provides a useful approximation of system resilience [17]. This broader understanding of resilience has informed a growing body of research on interdependent infrastructure systems. Although this research field has received increasing attention over the years [18–21], a review of studies indexed in Web of Science and Scopus shows that existing studies have largely focused on dependencies between electric power and water supply systems (for example, [22–25]), or between electric power and road systems (for example, [26–29]). Only a few studies have examined the resilience of rail systems while explicitly considering their dependence on electric power [30–42]. The dependency representations, cascading failure modelling approaches, resilience performance indicators, case studies, and disruption scenarios covered in these studies are summarised in Table 1.

A consistent feature across most of these studies is the representation of rail-power connection as a functional dependency, meaning that the operation of one system is dependent on the material outputs of another system [19]. The level of detail in representing this functional dependency varies across studies, ranging from broad associations between the metro network and the power grid [31,37,39,40] to explicit associations between individual track segments and their supplying traction substations [32,38] and, in some studies, further upstream links to the power distribution network [36,41]. Additionally, network model-based simulations are the dominant method for representing rail and power infrastructure and their dependencies, simulating cascading failures triggered by power outages, and evaluating system performance loss to assess resilience. The performance indicators used to assess URT resilience can be categorised into two groups: topological indicators characterise physical connectivity, such as network efficiency [35,37,39,42] and betweenness centrality [40], while operational indicators reflect service delivery outcomes, such as passenger delay [30,34], unmet travel demand [32], and the ratio of functional stations [41]. In terms of resilience modelling scope, about half of these studies incorporate post-disruption recovery modelling, albeit with varying levels of detail in how repair times, resource constraints, and restoration sequencing are considered [30–32,34–36,39]. The remaining studies adopt a robustness-oriented resilience assessment approach, assessing the system's ability to maintain functionality under disruption by measuring performance degradation. Finally, the case studies vary in scale, spanning a few stations [31,32], a single metro line [30,38,40], and multi-line urban rail networks [33,37,39,41].

**Table 1** Review of studies on URT or railway resilience considering electric power dependency

| Reference | Dependency type and detail | Dependency case study | Disruption scenario | Power supply redundancy | Cascading failure modelling | Modelling method | Performance indicator | Recovery modelling |
|---|---|---|---|---|---|---|---|---|
| [30] | Functional: Rectification Substation (RS) provides traction power. | One line of the Paris mass railway. | Three targeted traction power disruption scenarios: (1) one RS fails; (2) two non-adjacent RSs fail; (3) two adjacent RSs fail. | Considered by reducing train speeds. | Reduction of the speed of the trains around the failed RS. | Network model-based numerical simulation. | Operational: passenger delay and passenger load. | Considered staff travel time and RS repair time. |
| [31] | (1) Functional: metro operations rely on power supply. (2) Restoration: metro station closures delay power recovery by increasing maintainer travel time. | 10 stations of the New York subway. | One targeted disruption scenario caused by microgrid and backup generator failure. | ✗ | Power outage suspends the entire metro line; the number of working metro stations affects microgrid through a reward function. | Double-layer network model. | Operational: average number of working power buses and metro stations. | Apply Markov decision processes to model recovery dynamics. |
| [32] | (1) Stochastic failure propagation dependency. (2) Logic dependency between assets and sub-assets. (3) Asset utilisation dependency. (4) Resource input dependency. | 39 assets (inc. 17 stations) of the London Underground. | One reservoir failure-induced hypothetical flood scenario with 74 unique asset failure scenarios. | ✗ | Dependency type (1): probabilistic failure propagation; type (2): logic rule-based deterministic immediate effects; type (3): capacity reduction due to dependent asset inoperability; type (4): recovery constrained by competing resources. | Dynamic network flow model. | Operational: unmet travel demand. | Consider repair resources and time (deterministic). |
| [33] | Functional & geographic: closer assets have stronger interdependencies and higher cascading failure risk. | 5 lines of the Washington D.C. metro. | One hypothetical targeted power outage disruption scenario. | ✗ | Reduce the functionality of facilities within a specified radius. | Agent-based model. | Operational: functionality index of all facilities. | ✗ |
| [34] | Functional: track, train, and signalling rely on electricity; track and train also rely on signalling and telecoms. | Great Britain's southern rail network. | 34 targeted disruption scenarios, each involving the failure of one traction power supply point. | ✗ | Failure of one asset immediately triggers the failure of its dependent assets. | Interdependent network flow model. | Operational: train and passenger delay minutes. | Only a 60-min recovery time is applied. |
| [35] | Functional (focus on train operation): machine network (MN) and communication network (CN) rely on electricity network (EN); MN and EN also rely on CN. | High speed train in China. | Hypothetical scenario; details not specified. | ✗ | Failure of a node in one sub-network immediately causes failure of its connected nodes in other sub-networks. | Multi-layer network model. | Topological: network efficiency and node degree. Operational: power load and traffic load. | Sequential node recovery with same repair time per failed unit; one unit restored per time step. |
| [36] | Functional: metro relies on traction power supply. | An IEEE 33-node and an IEEE 123-node system. | Hypothetical targeted attack on the distribution power network. | Double-end feeding. | Metro track failure occurs only when both traction substations fail. | Multi-layer network model. | Operational: time cost of unsatisfied travel demand. | Considered repair time and repair crew sequencing. |

(continue Table 1)

| Reference | Dependency type and detail | Dependency case study | Disruption scenario | Power supply redundancy | Cascading failure modelling | Modelling method | Performance indicator | Recovery modelling |
|---|---|---|---|---|---|---|---|---|
| [37] | Functional: metro relies on municipal power supply. | 25 lines of the Beijing subway. | One historical earthquake scenario (with 1,000 Monte Carlo simulations). | ✗ | Extensive / complete damage to a substation immediately triggers failure of its connected metro stations. | Network model. | Topological: weighted network efficiency. | ✗ |
| [39] | Functional: metro relies on municipal power supply. | 25 lines of the Beijing subway. | One historical earthquake scenario (with 5,000 simulations). | ✗ | Extensive / complete damage to a substation immediately triggers failure of its connected metro stations. | Network model; deep-learning-based surrogate model. | Topological: weighted network efficiency. | Considered restoration sequence, duration, start time, and resources for damaged metro and power components. |
| [38] | Functional: metro stations rely on step-down substations; metro contact network lines rely on traction substations. | One line of the Xi'an metro. | Four hypothetical targeted attack scenarios: (1) attack one step-down substation; (2) attack two traction step-down hybrid substation, respectively; (3) attack one main substation. | ✗ | Substation failure immediately triggers failure of its connected metro components. | Double-layer network model. | Operational: passenger accessibility. | ✗ |
| [40] | Functional: metro relies on power supply. | One line of the Lyon metro. | Three hypothetical power outage scenarios: metro line operates at (1) reduced frequency; (2) with replacement service; (3) cessation of the entire line. | ✗ | ✗ (dependencies are not simulated but represented through an assumption that power outages cause failures in selected parts of the metro network). | Multi-layer network model. | Topological: betweenness centrality. Operational: average travel time. | ✗ |
| [41] | Functional: metro relies on traction power supply. | 9 lines of the metro in a metropolis in China. | Random and targeted attacks on the power network. | Double-end feeding. | Metro track failure occurs only when both traction substations fail. | Multi-layer network model. | Operational: ratio of functional stations and travel-load loss. | ✗ |
| [42] | Six theoretical coupling models. | The metro in a city in China. | Targeted attack: top 10% of nodes with the highest degree centrality are selected to fail. | ✗ | ✗ (not specified). | Multi-layer network model. | Topological: network efficiency. | ✗ |
| This study | (1) Functional: metro relies on traction power supply. (2) Geographic: nearby assets face similar hazard exposure. | 16 lines of the London Tube. | 6 hypothetical surface water flood scenarios: 30-, 100-, and 1,000-year flood risks under current and RCP8.5 climate scenarios. | Tested both: single-end and double-end feeding. | Single-end: track failure if either traction substation fails; double-end: track failure only if both traction substations fail. | Double-layer network model. | Operational: ratio of satisfied travel demand. | ✗ |

Despite these efforts, several limitations remain. First, redundancy in traction power supply mechanisms has rarely been considered. Most studies simulate cascading failures by assuming that the malfunction of a power-network component, such as a traction substation, immediately triggers the failure of the dependent metro components, reflecting a single-end feeding mechanism [33–35,38]. However, urban rail systems, particularly heavy metros that are predominantly Direct Current (DC)-electrified, commonly adopt a double-end feeding design for their traction power supply [43]. Under this mechanism, the track section between two adjacent traction substations is supplied from both ends and remains energised as long as at least one of the two substations is operational [44]. This provides inherent N-1 redundancy, whereby the loss of a single substation may reduce the local power margin and require reduced train speeds, but does not necessarily, by itself, disrupt train operations. Consequently, modelling traction substation cascading failures as directly propagating to dependent metro tracks may misrepresent the physical mechanism by which power-side disturbances propagate to the operational layer and thereby overstate the extent of cascading consequences. Second, existing studies remain limited in developing disruption scenarios that reflect real-world hazard conditions, as also highlighted in a recent review paper [45]. Disruption scenarios in resilience assessments are often abstract, relying on random or targeted component failures [30,34,36,38,41,42] or other simplistic illustrative scenarios [31–33,35,40], rather than explicitly considering location-specific asset exposure to real-world natural hazards such as flooding. As a result, studies accounting for flood-conditioned geographical dependencies between URT and power systems remain scarce.

Third, existing studies lack metrics that measure of the degree of flood-conditioned power dependency of URT; that is, the extent to which URT disruption under a given flood scenario is caused by power-side flood failures rather than by direct flooding of URT components. While URT clearly depends on traction power supply during normal operation, this baseline dependency does not reveal the extent to which power-side failures contribute to system performance loss under a specific flood scenario. As such uncertainty depends on the specific flood exposure, the degree of power dependency should be understood as a scenario-conditioned property of URT and systematically examined under varying extend of flood scenarios. This limitation also applies to resilience assessment of other dependent infrastructure systems, where hazard-conditioned degrees of dependency are seldom explicitly characterised [46]. Finally, data availability has been widely recognised as a major challenge in interdependent infrastructure resilience studies [19,21,46,47]. Detailed asset-level dependency data are often inaccessible due to data scarcity or sensitivity, which limits the feasibility and scalability of fine-grained dependency simulation in large-scale infrastructure resilience assessments. This long-standing challenge calls for methods capable of capturing dependency effects using readily available spatial information to produce meaningful insights.

To address these limitations, this study develops a novel network model-based methodology to quantify flood-conditioned power dependency of the URT network and its effects on resilience under varying flood scenarios. The methodology adopts the robustness-oriented approach to resilience assessment while also capturing network redundancy through the availability of alternative routes within the acceptable delay threshold. It comprises four modules: (a) constructing a double-layer network model that couples the URT operational layer with the URT traction power supply layer through geographical proximity; (b) developing flood disruption scenarios that capture both direct track flooding and cascading failures from traction substation flooding; (c) proposing a metric that measures the degree of flood-conditioned power dependency of URT; and (d) quantifying the disruption to passenger journeys (i.e., performance loss) solely attributable to cascading effects of traction substation flooding. To examine how traction power supply mechanisms and flood-failure criteria influence the assessment results, this study compares single-end and double-end feeding configurations and stress-tests different traction substation flood-failure thresholds. Based on these results, it further investigates the correlation between flood-conditioned power dependency and dependency-related performance loss of a URT network. The methodology is demonstrated through a case study on 16 lines of the London rail transit network under 30-, 100-, and 1,000-year surface water flood scenarios for both current and RCP8.5 climate conditions. This study contributes to resilience assessment of interdependent infrastructure systems in the following ways:

(a) It proposes, for the first time, the concept of hazard-conditioned system dependency, which reframes inter-system dependency from a static structural characteristic into a scenario-conditioned property. By accounting for the spatial pattern and severity of the hazard, this concept identifies which dependency links are "activated" and become disruptive under hazard conditions of interest, providing a more realistic basis for assessing cascading consequences and avoiding over- or under-estimation of dependency-related impacts in practical risk management.

(b) The simple metric designed for measuring the degree of flood-conditioned system dependency is formulated to accommodate different levels of data availability. It can be applied using readily available high-level spatial information when detailed asset-level dependency data are unavailable, while allowing more detailed dependency data to be incorporated where available. This flexibility supports scalable and transferable assessment of hazard-conditioned dependency across different interdependent infrastructure systems. Additionally, given its strong correlation with dependency-related performance loss across tested scenarios in the London case study, the metric can serve as a proxy for rapidly estimating dependency-induced cascading consequences without requiring complex performance simulations, thereby supporting strategic planning-level decision-making.

(c) This study demonstrates the significance of explicitly considering traction power supply redundancy in URT resilience assessment, which is often overlooked in current literature. By comparing single-end and double-end feeding configurations in the London case, this study shows that neglecting redundancy can substantially overestimate dependency-related cascading effects and thereby reshape resilience assessment results. The case study further suggests that DC-electrified urban rail networks can be very robust to flood-conditioned power cascading failures, with minimal substation flooding-attributable performance loss even under the most severe flood scenario examined.

The remainder of this paper is organised as follows. Section 2 introduces the methodology developed to quantify flood-conditioned power dependency of the URT network and assess its effects on resilience. Section 3 presents the London case study. Section 4 details the results, followed by a discussion in Section 5. Conclusions and suggestions for future work are given in Section 6.

## 2 Methodology

The overall research framework designed for quantifying flood-conditioned power dependency of URT and its correlation with system performance loss under flood scenarios is illustrated in Fig. 1. It begins with constructing a double-layer network model that represents the URT operational layer, the URT traction power supply layer, and the dependency links between them (Fig. 1(a)). Following this, flood scenarios are developed, using flood depth maps, to identify both directly flooded track segments and indirectly affected segments caused by cascading failures from flood-exposed traction substations (Fig. 1(b)). A metric is proposed to assess flood-conditioned power dependency of URT (Fig. 1(c)), and a robustness-oriented approach is used to quantify the flood-conditioned, dependency-induced performance loss (Fig. 1(d)). Finally, the correlation between flood-conditioned power dependency and dependency-induced performance loss is analysed across a range of flood scenarios. The remainder of this section introduces each module in detail.

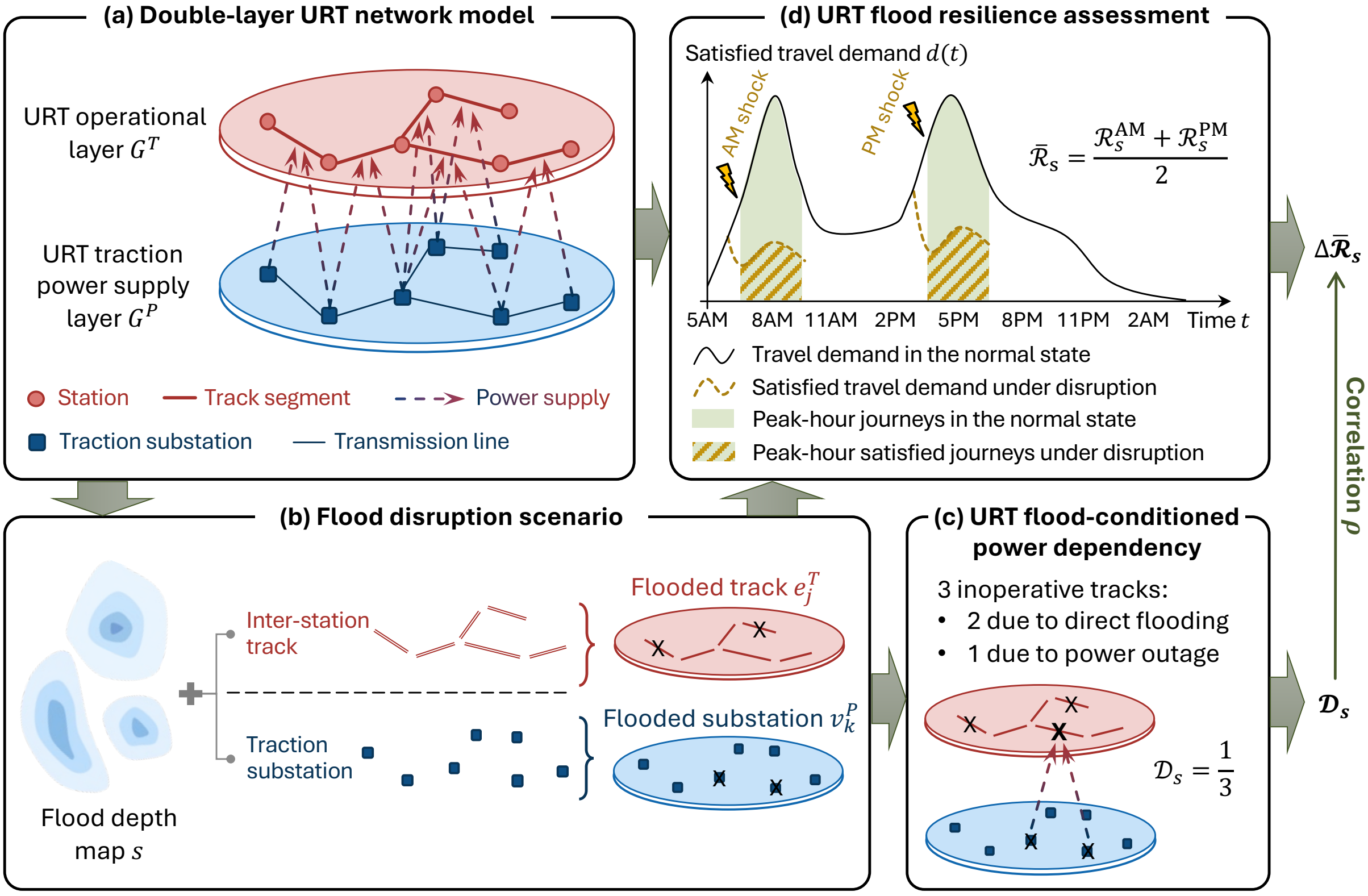

**Fig. 1** Research framework for quantifying flood-conditioned power dependency of URT and its correlation with performance loss

### *2.1 Double-layer URT network model*

The URT system is represented as a double-layer network model, denoted as $\mathcal{M} = (\mathcal{G}, \mathcal{C})$, where $\mathcal{G} = (G^T, G^P)$ consists of the URT operational layer $G^T$ and the traction power supply layer $G^P$, and $\mathcal{C}$ indicates the inter-layer coupling through dependency links between them.

The URT operational layer is modelled as a weighted graph $G^T = (V^T, E^T, W^T)$, where $V^T = \{v_i^T | i = 1,2,3, \dots, n\}$ is the set of $n$ nodes representing stations, $E^T = \{e_{ij}^T = (v_i^T, v_j^T) | i, j = 1,2,3, \dots, n; i \neq j\}$ is the set of edges representing track segments that connect two adjacent stations, and $W^T = \{\omega_{ij}^T | i, j = 1,2,3, \dots, n; i \neq j\}$ is the set of edge weights representing the travel time on each interstation track segment. The traction power supply layer is represented as $G^P = (V^P, E^P)$, where $V^P = \{v_k^P | k = 1,2,3, \dots, p\}$ is the set of $p$ nodes representing traction substations, and $E^P = \{e_{kl}^P = (v_k^P, v_l^P) | k, l =$

$1,2,3, \dots, p; k \neq l\}$ is the set of edges representing transmission lines that connect two adjacent traction substations.

The inter-layer coupling $\mathcal{C}$ is formulated as a binary bipartite incidence matrix that encodes the functional dependency between traction substations $v_k^P$ and URT track segments $e_{ij}^T$. A non-zero entry indicates that a track segment is supplied by a given traction substation, defining a directed dependency link from the traction power supply layer to the operational layer:

$$\mathcal{C} \in \{0,1\}^{p \times m}, \qquad \mathcal{C}_{k,ij} = \begin{cases} 1 & \text{if substation } v_k^P \text{ feeds track segment } e_{ij}^T \\ 0 & \text{otherwise} \end{cases} \tag{1}$$

where $k = 1,2,3, \dots, p$, $e_{ij}^T \in E^T$, and $m = |E^T|$ is the number of track-segment edges. As explained in the introduction and illustrated in Fig. 2, DC-electrified URT usually adopt a double-end feeding design for their traction power supply [43], whereby each track segment between two adjacent traction substations is supplied from both ends. Therefore, the inter-layer coupling matrix is constrained such that each track segment is connected to exactly two traction substations:

$$\sum_{k=1}^{p} \mathcal{C}_{ij} = 2 \tag{2}$$

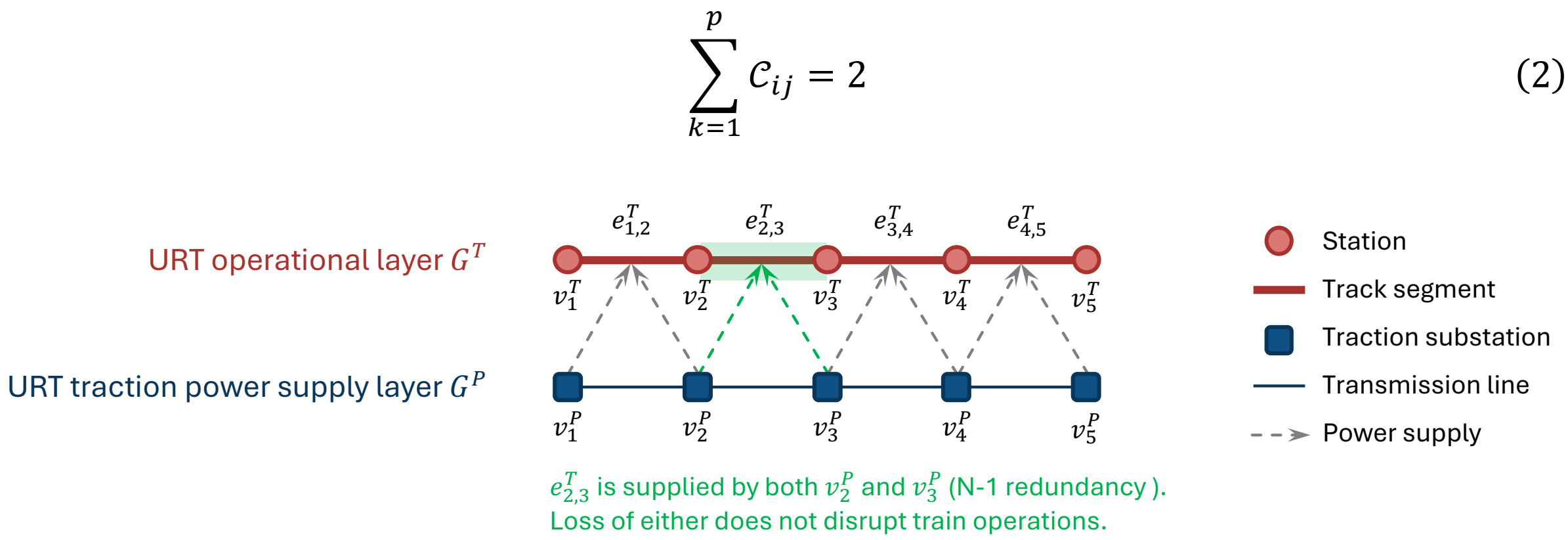

**Fig. 2** Illustration of double-end feeding in a DC-electrified URT traction power supply system

### *2.2 Flood scenarios*

Infrastructure resilience analytics follows a stress-testing approach, assessing how a system performs under potentially disruptive scenarios [48,49]. For flood scenarios, flood-depth maps are commonly used to identify assets that are exposed to inundation and may consequently lead to functional disruption [50,51]. Inundated URT stations prevent passengers from entering or exiting the system, and flooded tracks disrupt train operations [52]. While both stations and track segments are susceptible to flood-induced functional failure, this study focuses exclusively on disruptions caused by track flooding for three reasons: (a) outdoor tracks are an important contributor to weather-related URT service disruptions but remain underexamined in the literature [52,53]; (b) accounting for station flooding would hinder the isolation of performance loss attributable specifically to the dependency between URT operations and traction power supply; and (c) station operation relies primarily on auxiliary power supply, which constitutes another layer of power dependency beyond the scope of this study.

### *2.2.1 Identification of flood-exposed and inoperative assets*

Let $\phi_s(\ell)$ indicate the flood depth at asset $\ell$ under flood scenario $s$. In this study, outdoor track segments are examined for identifying direct track inundation, while traction substations are considered to capture flood-induced power outages and their cascading effects on dependent track segments. Let $\xi_s(\ell)$ indicate the binary functionality of asset $\ell$ under flood scenario $s$ resulting from direct inundation, where $\xi_s(\ell) = 1$ indicates that the asset remains functional and $\xi_s(\ell) = 0$ indicates that it is directly flooded and thereby inoperative. An asset is considered inoperative when the flood depth reaches or exceeds its specified threshold:

$$\xi_s\left(e_{ij}^T\right) = \begin{cases} 0 & \text{if } \phi_s\left(e_{ij}^T\right) \geq \phi_{e^T}^* \\ 1 & \text{otherwise} \end{cases} \tag{3}$$

$$\xi_s\left(v_k^P\right) = \begin{cases} 0 & \text{if } \phi_s\left(v_k^P\right) \geq \phi_{v^P}^* \\ 1 & \text{otherwise} \end{cases} \tag{4}$$

where parameters $\phi_{e^T}^*$ and $\phi_{v^P}^*$ refer to the flood-depth thresholds for track segments and traction substations, respectively.

### *2.2.2 Failure propagation*

In addition to direct flood-induced failures, a track segment may become inoperative because of traction power loss caused by the flooding of its dependent substations. Let $\xi_s^{pwr}(e_{ij}^T)$ denote the binary functionality of track segment $e_{ij}^T$ under flood scenario $s$ arising from dependency-induced traction power loss. Specifically, $\xi_s^{pwr}\left(e_{ij}^T\right) = 1$ indicates that the track segment remains adequately supplied by its dependent traction substations, while $\xi_s^{pwr}\left(e_{ij}^T\right) = 0$ indicates that it becomes inoperative because flooding has disabled the substations on which it relies. Under the typical double-end feeding mechanism, $\xi_s^{pwr}\left(e_{ij}^T\right) = 0$ only when both supplying substations are disrupted. To further assess the effect of traction power redundancy, a non-redundant feeding mechanism (referred to earlier as single-end feeding) is also considered, where the failure of either supplying substation renders the track segment inoperative. The dependency-induced functional state is therefore defined as:

$$\xi_s^{pwr}\left(e_{ij}^T\right) = \begin{cases} 1 - \prod\limits_{k:\mathcal{C}_{k,ij}=1} \left(1 - \xi_s\left(v_k^P\right)\right), & \text{under double} - \text{end feeding} \\ \prod\limits_{k:\mathcal{C}_{k,ij}=1} \xi_s\left(v_k^P\right), & \text{under non} - \text{redundant feeding assumption} \end{cases} \tag{5}$$

The final operability of a track segment $\mathcal{Y}_s\left(e_{ij}^T\right) \in \{0,1\}$ is determined jointly by direct flooding and dependency-induced traction power loss:

$$\mathcal{Y}_s\left(e_{ij}^T\right) = \xi_s\left(e_{ij}^T\right) \cdot \xi_s^{\text{pwr}}\left(e_{ij}^T\right) \tag{6}$$

Thus, a track segment remains operational only if it is neither directly inundated nor affected by the flooding of its supplying substations. Under each flood scenario $s$, edges representing inoperative track segments are disabled in the network model to simulate service disruptions.

### *2.3 Flood-conditioned power dependency of URT*

Building on the distinction between direct flood-induced failures and dependency-induced traction power loss, a metric is proposed to characterise the flood-conditioned power dependency of a URT network. Rather than treating power dependency as a fixed system characteristic, the metric represents it as a scenario-conditioned property that varies with the extent and spatial distribution of flood exposure. Let $\mathcal{D}_s$ denote the flood-conditioned power dependency of URT under scenario $s$, which is defined as the proportion of inoperative track segments whose loss of functionality is attributable solely to traction power disruption caused by substation flooding:

$$\mathcal{D}_s = \frac{\left|\left\{e_{ij}^T \in E^T : \xi_s\left(e_{ij}^T\right) = 1 \wedge \xi_s^{pwr}\left(e_{ij}^T\right) = 0\right\}\right|}{\left|\left\{e_{ij}^T \in E^T : \psi_s\left(e_{ij}^T\right) = 0\right\}\right|} \tag{7}$$

By definition, $\mathcal{D}_s \in [0,1]$. $\mathcal{D}_s = 1$ indicates that all inoperative track segments result solely from traction power loss, with none directly inundated. Conversely, $\mathcal{D}_s = 0$ indicates that no track segment becomes inoperative solely because of traction-substation flooding: all inoperative segments are directly inundated, although some may also lose traction power. Where a segment is affected by both direct inundation and power loss, its inoperability is attributed to direct inundation to avoid double counting, since it would already be inoperative regardless of the power-supply state. The indicator therefore captures only the additional disruption caused by substation flooding. Any delay in track recovery due to slower substation restoration is beyond the scope of this study because the detailed recovery process is not modelled, as explained later in Section 2.4.2.

### *2.4 URT flood resilience assessment*

#### *2.4.1 Performance indicator*

Infrastructure resilience is commonly quantified by the loss of system performance following disruption [16]. This requires a practically meaningful performance indicator that reflects the network's ability to maintain service under disrupted conditions. In this study, URT performance is indicated by hourly satisfied travel demand, which is defined as the number of planned journeys per hour that can still be completed under a disruption. This indicator is practically meaningful because it not only captures the network's ability to provide alternative routes from a physical connectivity perspective, but also directly reflects the operational performance in terms of the extent to which normal passenger demand can continue to be accommodated [52].

Following a commonly used assumption in transport studies, passengers are assumed to travel along the path with the shortest total travel time between their origin and destination (OD) stations, where the total travel time accounts for train running time between stations, dwell time at intermediate stations, and transfer time at interchange stations. A journey is considered satisfied under a disruption when (a) the OD pair remains connected and (b) the increase in shortest travel time relative to the undisrupted network does not exceed the acceptable delay threshold $\Delta\lambda^*$. The delay threshold serves as a simplified proxy for passenger responses to excessive increases in travel time, such as switching to alternative modes or cancelling the trip. In either case, these affected journeys should be considered a loss of URT operational performance.

Let $\sigma_s(v_O^T, v_D^T, t)$ denote the journey-satisfaction indicator for travel from origin station $v_O^T$ to destination station $v_D^T$ at hour $t$ under flood scenario $s$. A value of 1 indicates that the journey remains satisfied, whereas a value of 0 indicates that it is disrupted. According to the satisfaction criteria specified above, the indicator is defined as follows:

$$\sigma_s(v_O^T, v_D^T, t) = \begin{cases} 1 & \text{if } \lambda_s(v_O^T, v_D^T, t) < \infty \text{ and } \lambda_s(v_O^T, v_D^T, t) - \lambda_0(v_O^T, v_D^T) \leq \Delta\lambda^* \\ 0 & \text{otherwise} \end{cases} \tag{8}$$

Here, $\lambda_0(v_O^T, v_D^T)$ refers to the baseline shortest travel time between the OD pair $(v_O^T, v_D^T)$ in the undisrupted URT network, while $\lambda_s(v_O^T, v_D^T, t)$ is the shortest travel time in the disrupted URT network at hour $t$ under scenario $s$. When no feasible path exists between the origin and destination stations, $\lambda_s(v_O^T, v_D^T, t)$ is set to infinity. The acceptable delay threshold $\Delta\lambda^*$ is set to 30 minutes, following the threshold adopted and justified in previous studies [5,52,54]. The satisfied travel demand $Q_s(t)$ at hour $t$ under scenario $s$ is then calculated as:

$$Q_s(t) = \sum_{(O,D)} d_{OD,s}(t) \cdot \sigma_s(v_O^T, v_D^T, t) \tag{9}$$

where $d_{OD,s}(t)$ is the number of planned journeys from station $v_O^T$ to station $v_D^T$ during hour $t$.

*2.4.2 Robustness-oriented resilience metric*

Resilience assessment captures the temporal dynamics of a system, with its most comprehensive form evaluating the cumulative loss of performance from disruption onset to recovery completion [55]. In this study, however, recovery is not explicitly modelled. Instead, the assessment focuses on the system's ability to withstand flood-induced disruption and therefore uses robustness as a proxy for resilience, as adopted in many of the studies reviewed in Table 1. It also captures network redundancy by allowing passenger journeys to be rerouted within the acceptable delay threshold when the original path is disrupted. This approach is appropriate for network-level flood-risk screening undertaken in this study, whereas recovery modelling would be more meaningful for realistic, event-specific flood disruption scenarios, which are typically localised and do not disrupt the entire URT network simultaneously.

As such, the robustness-oriented resilience metric $\mathcal{R}_s^f$ is defined as the ratio of total satisfied journeys to total planned journeys over a period $\mathcal{T}_f$:

$$\mathcal{R}_s^f = \frac{\int_{t_0^p}^{t_0^f + \mathcal{T}_f} Q_s(t)dt}{\int_{t_0^f}^{t_0^f + \mathcal{T}_f} \sum_{(O,D)} d_{OD,s}(t)\, dt} \tag{10}$$

where $t_0$ indicates the hour of disruption onset. Evaluating disrupted journeys over an entire day would implicitly assume that the network remains in its initial post-disruption state throughout the day, without accounting for operational response or recovery actions. Therefore, the assessment is restricted to the morning and afternoon peak periods, which provide a more meaningful basis for evaluating the immediate effects of disruption under high-demand conditions. Eq. (10) is applied to morning and afternoon peak periods to obtain $\mathcal{R}_s^{\mathrm{AM}}$ and $\mathcal{R}_s^{\mathrm{PM}}$, respectively, as presented in Fig. 1(d). The overall resilience index for scenario $s$ is then defined as the mean of the two values:

$$\bar{\mathcal{R}}_s = \frac{1}{|\mathcal{F}|} \sum_{f \in \mathcal{F}} \mathcal{R}_s^f, \qquad \mathcal{F} \in \{\mathrm{AM}, \mathrm{PM}\} \tag{11}$$

To explicitly assess the effect of power dependency-induced service disruptions, $\bar{\mathcal{R}}_s$ is compared with the corresponding resilience index obtained when only disruptions caused by direct track flooding are considered, effectively assuming a fully flood-resilient traction power supply subsystem. The resulting difference, denoted by $\Delta\bar{\mathcal{R}}_s$, captures the performance loss attributable solely to traction power disruption caused by substation flooding and is used later to investigate its correlation with the flood-conditioned power dependency index $\mathcal{D}_s$.

### *2.5 Correlation analysis*

Spearman's rank correlation coefficient, also referred to as Spearman's $\rho$ [56], is adopted to assess the correlation between the flood-conditioned power dependency index $\mathcal{D}_s$ and the corresponding dependency-induced performance loss $\Delta\bar{\mathcal{R}}_s$ for each flood scenario $s$ across different modelling configurations. These modelling configurations vary in terms of the traction power feeding mechanism and the flood-depth threshold $\phi_{v^P}^*$ used to determine substation failure, with each configuration yielding a corresponding pair of $\mathcal{D}_s$ and $\Delta\bar{\mathcal{R}}_s$ values for a given flood scenario. The specific configurations considered in the case study are described later in Section 3.3. Spearman's $\rho$ ranges from $-1$ to 1, with positive and negative values indicating positive and negative monotonic correlations, respectively. Values closer to either extreme indicating a stronger correlation.

## 3 Case and data description

### *3.1 The London URT*

The London URT system is a highly complex network that serves millions of passenger journeys every day [57] while remaining exposed to substantial surface water flood risk [58]. It is therefore selected as the case study for demonstrating the proposed methodology. The system comprises several modes, including the London Underground (LU), London Overground (LO), Docklands Light Railway (DLR), Elizabeth line, and London Trams. As shown in Fig. 3, the case study includes all 11 LU lines, two LO lines, and all three DLR lines, which operate on 630 V or 750 V DC third- or fourth-rail electrification [59–61] and are modelled using a double-end feeding mechanism. Accordingly, the operational layer of the London URT network model consists of 343 station nodes and 401 edges representing interstation track segments. The network topology is derived from the official London Tube map [62], while station locations and track geometries are obtained from OpenStreetMap.

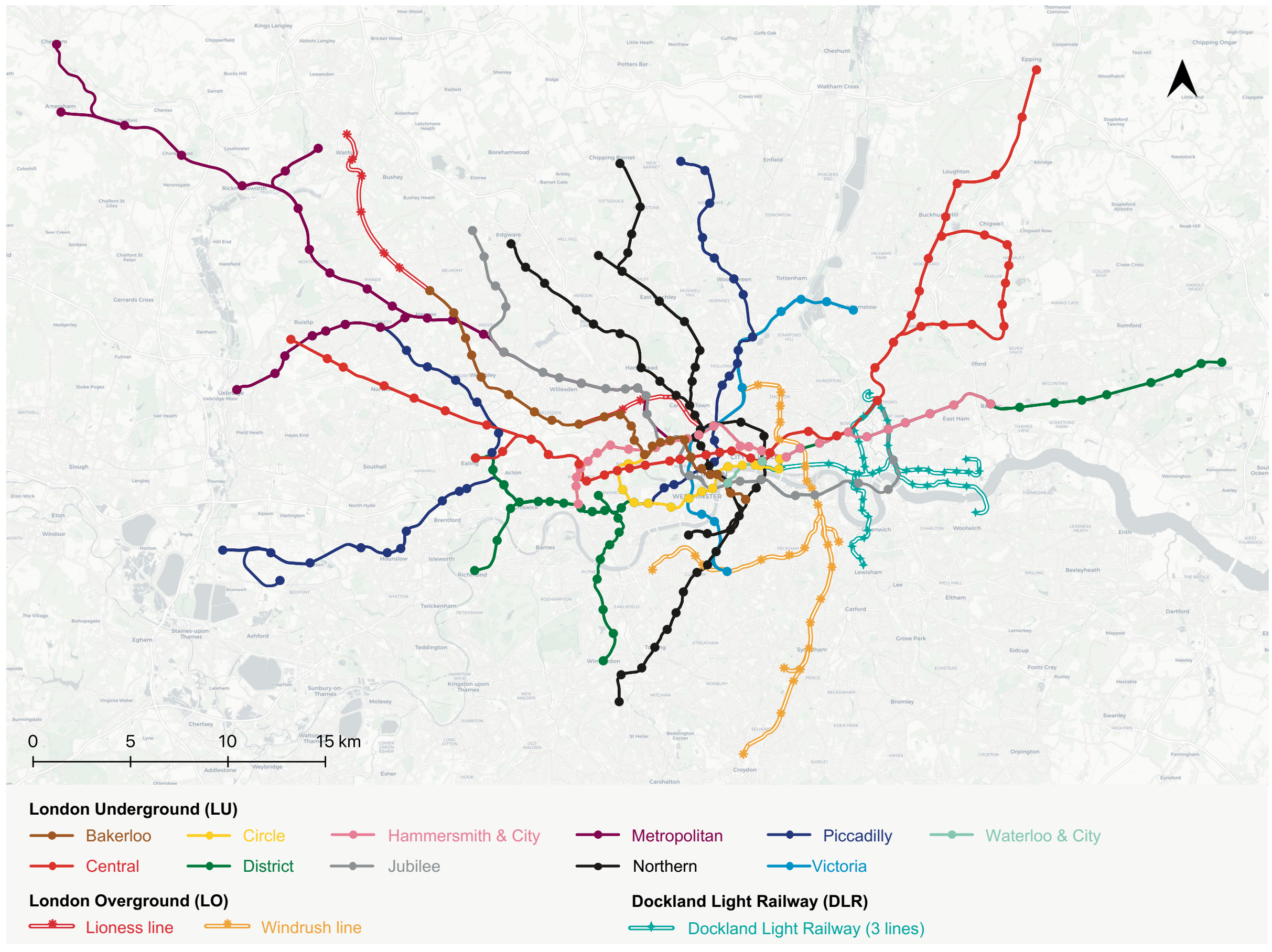

**Fig. 3** The sixteen lines of the London URT covered in the case study

The traction power supply layer focuses on traction substations, while upstream bulk supply points (132/22 kV) and high-voltage distribution network are excluded, as the upstream network is designed with a high degree of redundancy and historical evidence suggests that the loss of a single bulk supply point does not necessarily disrupt train services [63]. Taking the LU as an example, electricity is drawn from the National Grid at bulk supply points and distributed through 22 kV and 11 kV networks [44]. Traction substations then convert the 22 kV or 11 kV AC supply to DC through transformer-rectifier units for conductor-rail operation [60]. Because traction substation locations are security-sensitive and not publicly disclosed, a proxy dataset is constructed using publicly available UK Power Networks (UKPN) datasets, which have been adopted in previous studies to represent substation locations in London URT and UK railway case studies [32,34]. Candidate substations are derived from the UKPN 22 kV primary substation dataset [64] and 11 kV secondary substation dataset [65], supplemented by OpenStreetMap queries, and spatially screened first against a 150 m buffer around the track alignments. Following this, the nearest identified substation to each URT station was provisionally assigned [66] and subsequently reviewed using Google Street View. Where a more plausible traction substation location is identified, it replaces the provisional assignment. This procedure resulted in 334 proxy traction substation nodes, of which 328 are associated with a single URT station and six serve two URT stations. These proxy locations provide a reproducible spatial representation of the traction power supply layer for flood exposure and dependency analysis, while allowing authoritative asset data to be incorporated directly when available.

Passenger travel demand data are obtained from Transport for London (TfL). This study uses hourly station-to-station demand data for a typical weekday in 2024, provided by TfL upon request because the network-wide OD dataset for 2024 is not publicly available through its online data portal [67]. Fig. 4 shows the hourly demand served by the 16 lines included in the case study. Across 46,912 OD pairs, the morning peak period (7–9 AM) comprises 863,450 journeys, while the afternoon peak period (4–6 PM) comprises 1,004,042 journeys.

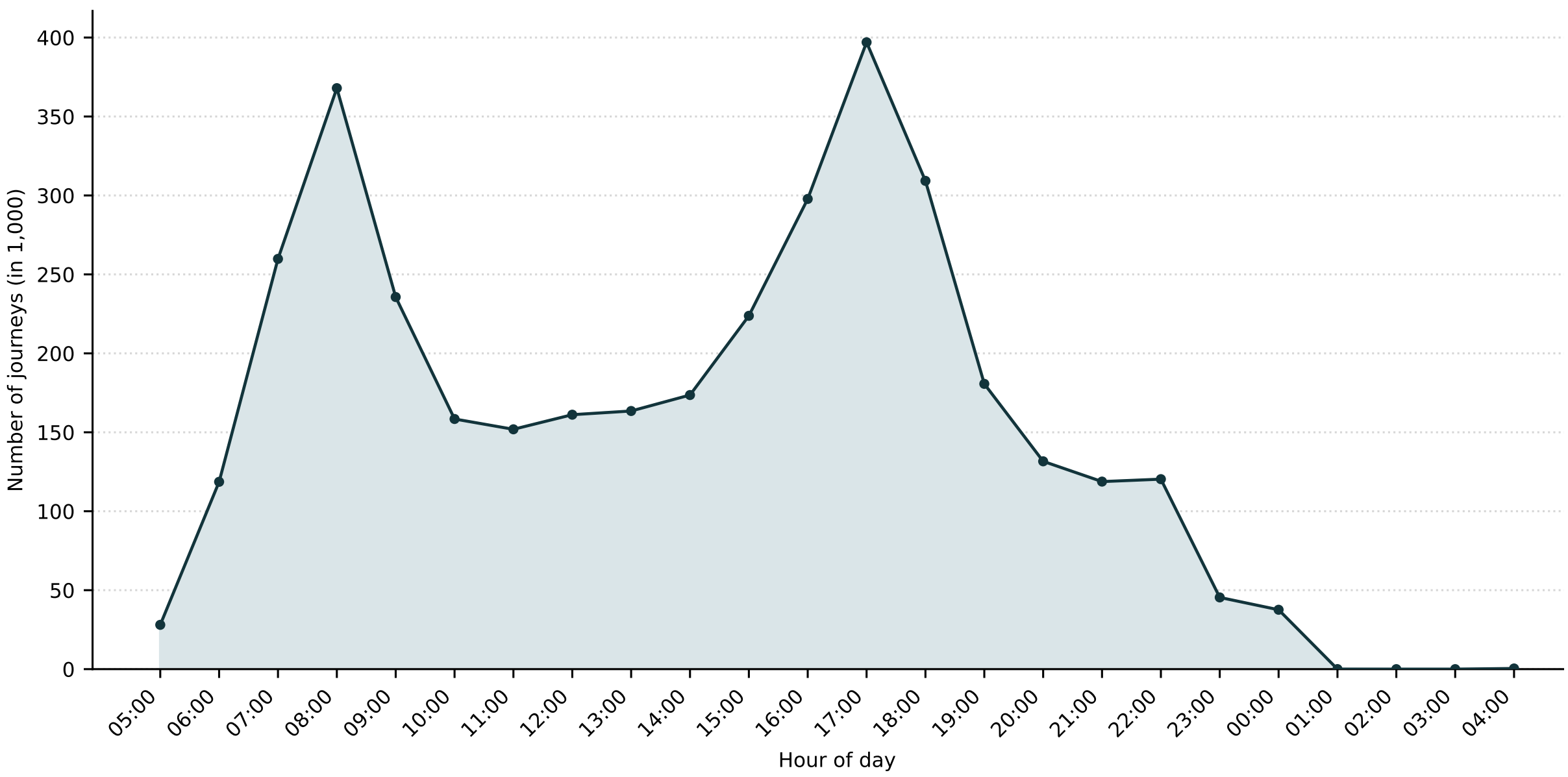

**Fig. 4** Hourly travel demand served by the 16 London URT lines on a typical weekday in 2024

### *3.2 Flood scenarios*

Using surface water flood risk maps published by the UK Environment Agency, this study examines six scenarios covering 30-, 100-, and 1,000-year return periods under current [68] and RCP8.5 [69] climate conditions. The maps have a horizontal grid resolution of 2 m, enabling location-specific flood exposure to be identified for track segments and traction substations. Each grid cell is classified into one of six flood depths: 0 m, 0.2 m, 0.3 m, 0.6 m, 0.9 m, and 1.2 m.

In this case study, the flood-depth threshold for determining inoperability of track segments, denoted by $\phi^*_{e^T}$ in Section 2.2.1, is set to 0.30 m. This value reflects engineering considerations, including rail-head height and signalling failure threshold, and has been adopted in previous studies following consultation with TfL professionals [5,52,54]. For each track segment, a 1 m buffer is created around its GIS geometry and overlaid with each flood-depth map. A track segment is considered exposed where a continuous flooded area within the buffer exceeds 127 $m^2$, a threshold representing half the average platform-track length [5]. Among all flood-depth areas satisfying this condition, the maximum depth is assigned to the track segment. The segment is then classified as inoperative if the assigned flood depth is equal to or greater than 0.30 m.

Regarding the flood-depth threshold for determining inoperability of traction substations, denoted by $\phi^*_{v^P}$, values adopted in previous studies vary widely, from approximately 0.20 m to 1.50 m [70–77].

This variation likely reflects differences in substation design, equipment layout, flood protection, and local site conditions, making it impossible to justify a single threshold for all assets. Therefore, this study examines four thresholds separately, including 0.3 m, 0.6 m, 0.9 m, and 1.2 m. This approach captures uncertainty in substation flood vulnerability and helps assess the level of flood protection required to reduce substation flooding-induced service disruption. To determine the flood depth at each traction substation, a circular buffer with a 5 m radius is created around the substation point and overlaid with the flood map. A substation is considered exposed when flooded areas cover at least 40% of the buffer [5]. If at least 70% of the flooded area within the buffer has the same flood depth, that depth is assigned to the substation [5]. Otherwise, an area-weighted mean flood depth is calculated and rounded down to the highest of the four threshold values that does not exceed the calculated mean [5].

### *3.3 Simulations*

The network modelling process, including the simulation of network disruptions and journey satisfaction, is implemented in Python using the NetworkX library. In total, 54 distinct simulations are conducted. Six baseline simulations considered service disruption caused solely by direct track flooding, one for each flood scenario. A further 48 simulations considered both direct track flooding and dependency-related disruption caused by traction substation flooding, covering all combinations of six flood scenarios, four substation flood depth thresholds, and two traction power supply configurations. For each combined-disruption simulation, the corresponding direct flooding baseline is used to quantify the dependency-induced performance loss $\Delta\bar{\mathcal{R}}_s$.

## 4 Results

### *4.1 Flood-exposed assets and flood-conditioned power dependency index*

Following the procedure described in Section 3.2, Table 2 presents the number of track segments directly flooded within flood zones and those that may be indirectly disrupted by cascading failures caused by flood-exposed substations under six surface water flood scenarios, with Fig. 5 providing a visual statistical comparison across scenarios and flood-depth categories and Fig. 6 indicating the locations of affected tracks. The results show that the number of directly flooded track segments increases with flood severity, from 41 under the Baseline-30-year scenario to 129 under the Baseline-1,000-year scenario, and from 62 under the RCP8.5-30-year scenario to 144 under the RCP8.5-1,000-year scenario. A similar pattern is observed for indirectly affected track segments connected to one flood-exposed substation. Under the most severe scenarios, 102 and 99 track segments are connected to a single flood-exposed substation under the Baseline-1,000-year and RCP8.5-1,000-year scenarios, respectively, exposing them to potentially significant cascading disruption under a single-end feeding configuration. The slightly lower number under RCP8.5-1,000-year is due to the greater overlap between direct flooding and substation-related exposure: 50 of these track segments are also directly flooded, compared with 32 under Baseline-1,000-year. As these segments are already counted as directly affected, they do not appear as additional indirect impacts. By contrast, the number of track segments connected to two flood-exposed substations is very small across all scenarios, with most exposed to areas with relatively low flood depths, reaching a maximum of 12 under the RCP8.5-1,000-year scenario. This suggests potential minimal cascading effects under a double-end feeding configuration (as illustrated later in Section 4.2).

**Table 2** Number of London URT track segments directly and indirectly affected under flood scenarios

| Flood scenario | Flood depth (m) | Number of tracks directly affected by flooding | | Number of tracks connected to one flood-exposed substation | | Number of tracks connected to two flood-exposed substations | |
|---|---|---|---|---|---|---|---|
| | | Per depth | Total (ratio) | Per depth | Total (ratio) | Per depth | Total (ratio) |
| Baseline: 30-year | 0.3 – 0.6 | 25 | 41 (10%) | 9 | 13 (3%) | 1 | 1 (0.2%) |
| | 0.6 – 0.9 | 6 | | 1 | | | |
| | 0.9 – 1.2 | 4 | | 2 | | | |
| | ≥ 1.2 | 6 | | 1 | | | |
| Baseline: 100-year | 0.3 – 0.6 | 30 | 74 (18%) | 9 | 16 (4%) | 1 | 1 (0.2%) |
| | 0.6 – 0.9 | 19 | | 4 | | | |
| | 0.9 – 1.2 | 11 | | | | | |
| | ≥ 1.2 | 14 | | 3 | | | |
| Baseline: 1,000-year | 0.3 – 0.6 | 28 | 129 (32%) | 42 | 70 (17%) | 2 | 2 (0.5%) |
| | 0.6 – 0.9 | 28 | | 20 | | | |
| | 0.9 – 1.2 | 26 | | 5 | | | |
| | ≥ 1.2 | 47 | | 3 | | | |
| RCP8.5: 30-year | 0.3 – 0.6 | 25 | 62 (15%) | 12 | 17 (4%) | 1 | 1 (0.2%) |
| | 0.6 – 0.9 | 22 | | 2 | | | |
| | 0.9 – 1.2 | 5 | | | | | |
| | ≥ 1.2 | 10 | | 3 | | | |
| RCP8.5: 100-year | 0.3 – 0.6 | 39 | 102 (25%) | 18 | 28 (7%) | | 0 |
| | 0.6 – 0.9 | 21 | | 6 | | | |
| | 0.9 – 1.2 | 13 | | 2 | | | |
| | ≥ 1.2 | 29 | | 2 | | | |
| RCP8.5: 1,000-year | 0.3 – 0.6 | 26 | 144 (36%) | 64 | 99 (25%) | 11 | 12 (3%) |
| | 0.6 – 0.9 | 22 | | 25 | | 1 | |
| | 0.9 – 1.2 | 22 | | 5 | | | |
| | ≥ 1.2 | 74 | | 5 | | | |

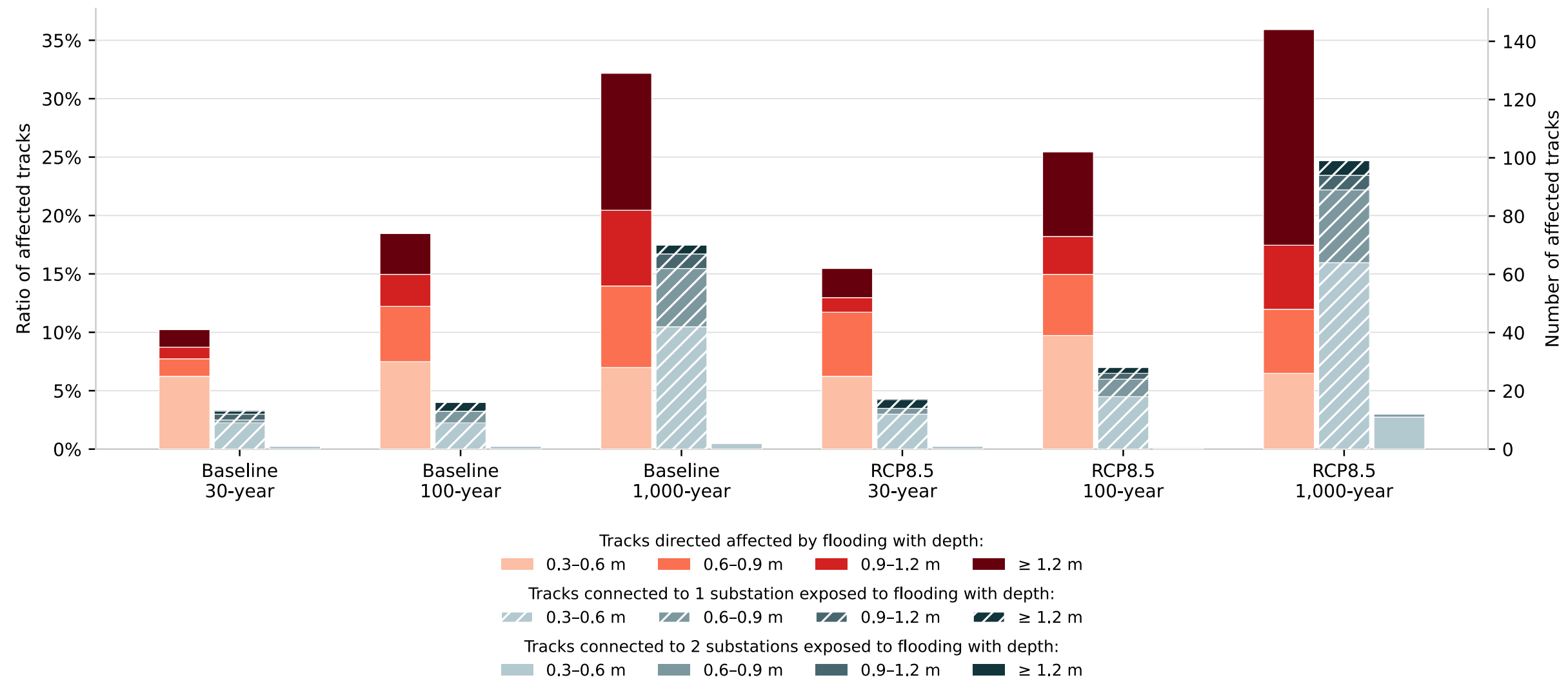

**Fig. 5** Number of London URT track segments directly and indirectly affected by flooding

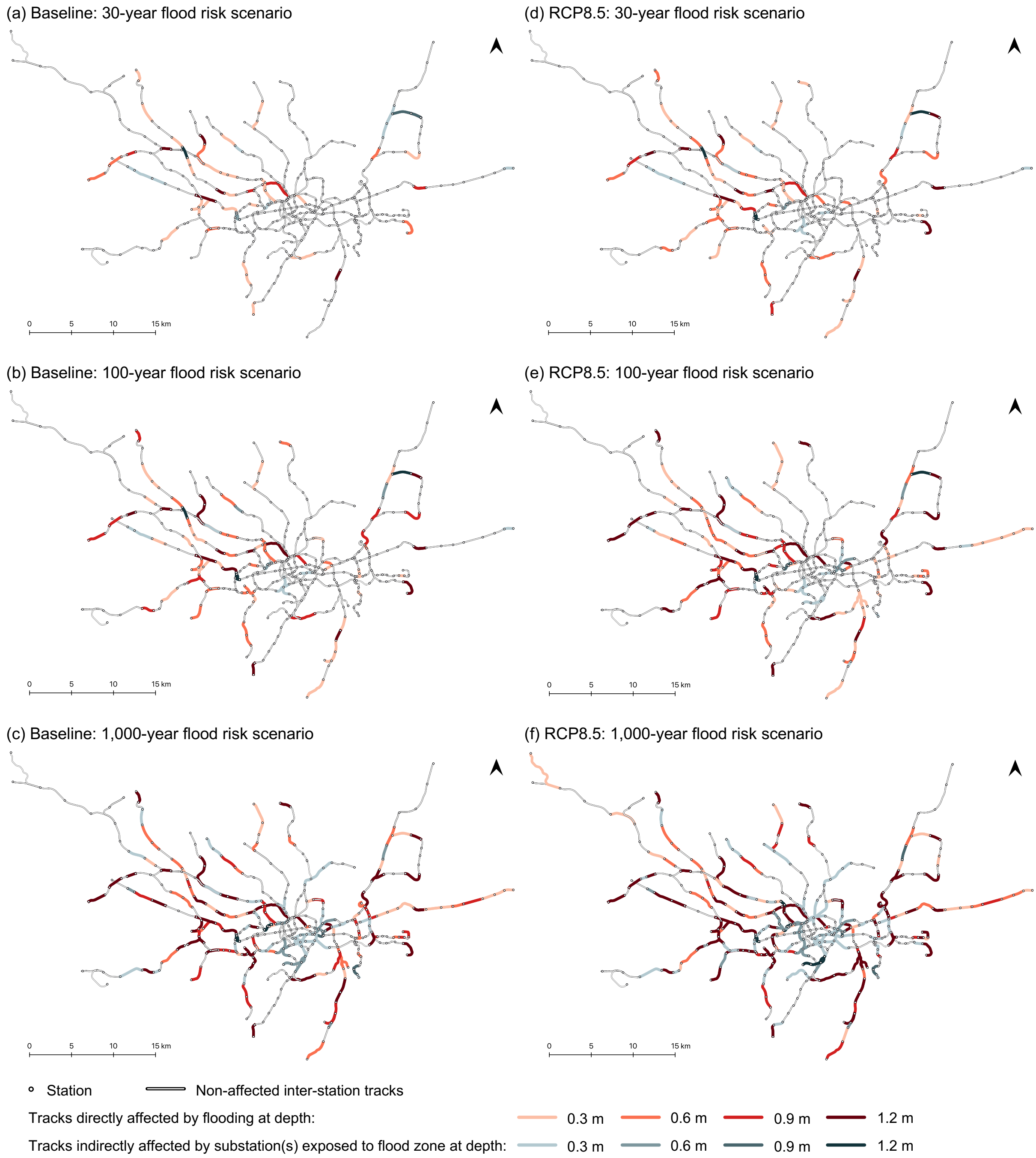

**Fig. 6** Locations of London URT track segments directly and indirectly affected by floods

Using Eq. (7), Fig. 7 presents the flood-conditioned power dependency index of the London URT across six surface water flood scenarios under single-end and double-end feeding configurations, each evaluated at four substation flood depth thresholds. Under double-end feeding, the dependency index remains close to zero in most cases, reflecting the limited number of track segments for which both feeding substations are flood-exposed, as shown in Table 2. By contrast, under single-end feeding, the index is greatly sensitive to the assumed substation flood-depth threshold. It is highest at the 0.3 m threshold and drops markedly at 0.6 m across all scenarios, indicating that much of the dependency-related exposure is associated with substations exposed to relatively shallow flooding. At thresholds above 0.9 m, the index remains low even under the most severe 1,000-year flood scenarios, suggesting that 0.9 m could represent an effective flood-tolerance threshold for limiting power dependency-related cascading disruption. The 100-year scenarios show lower dependency values than the 30-year scenarios in several cases, despite their greater flood severity, due to the larger overlap between direct flooding and substation-related exposure. This further highlights the uncertainty and non-linearity of power-related cascading failures, where more severe flooding does not necessarily translate into greater dependency-related disruption.

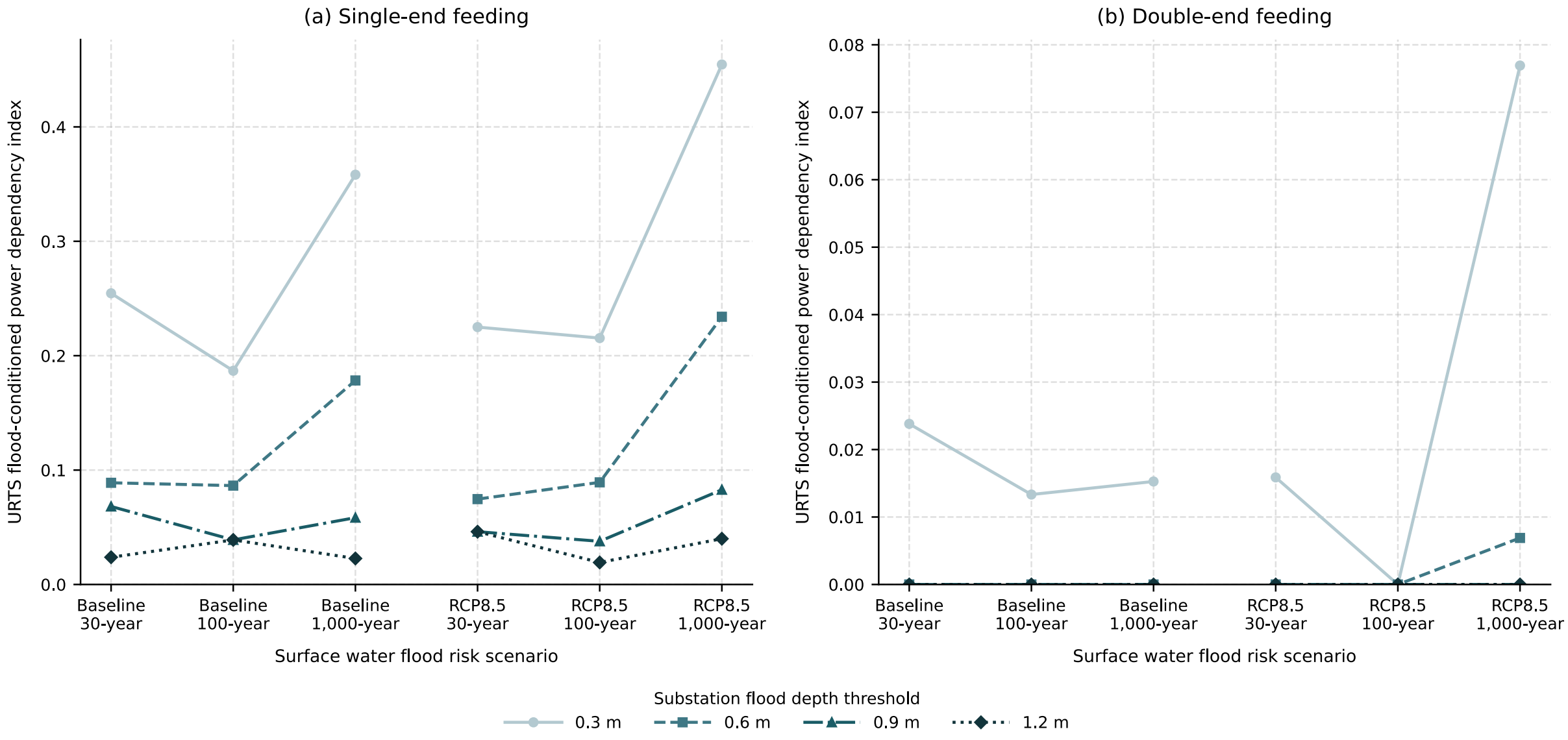

**Fig. 7** London URT flood-conditioned power dependency index

### *4.2 Peak-hour journey impacts under substation cascading effects*

Building on the above asset flood exposure results, this section quantifies the additional journey disruption solely caused by substation flood cascading effects beyond direct flood impacts. Fig. 8 presents the number of additional disrupted journeys and their share of total disrupted journeys (i.e., $\Delta\bar{\mathcal{R}}_S$ proposed in Section 2.4.2), averaged over the morning (i.e., 7-9 AM) and afternoon (i.e., 4-6 PM) peak hours. Under single-end feeding, these cascading effects substantially increase disrupted journeys across most flood scenarios, particularly at lower substation flood-depth thresholds. The largest additional impact occurs under the RCP8.5-1,000-year scenario at the 0.3 m threshold, with 461,066 additional disrupted journeys, accounting for 49.3% of total disrupted journeys in this setting. A similarly large contribution is observed under the Baseline-1,000-year scenario, where 371,344 additional disrupted journeys account for 39.8% of the total. These impacts decline sharply as the assumed substation flood-depth threshold increases. For instance, under RCP8.5-1,000-year, the additional disruption falls to 237,293 (25.4%) journeys at 0.6 m and further to 59,797 (6.4%) and 50,551 (5.4%) journeys at 0.9 m and 1.2 m, respectively. For the 30-year and 100-year scenarios, substation flood cascading effects also lead to noticeable disrupted journeys, although their share of total disrupted journeys remains relatively limited across most thresholds.

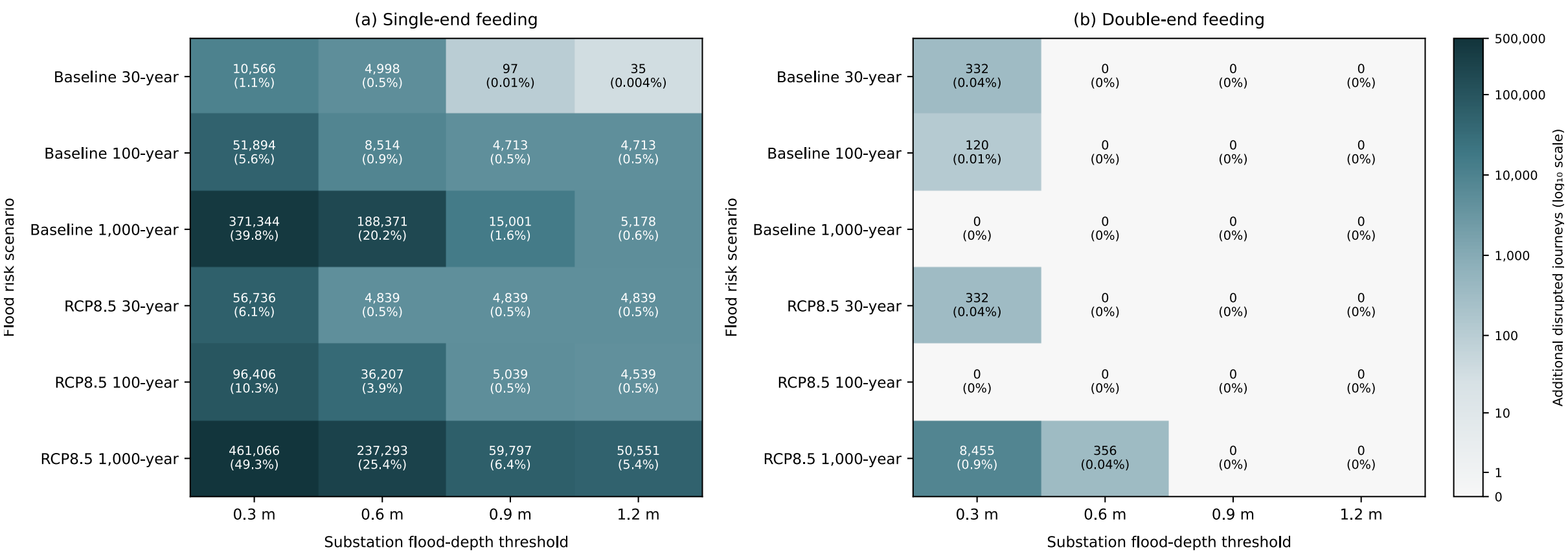

**Fig. 8** Additional disrupted journeys from substation cascading effects and their share of total disruptions (averaged over morning and afternoon peaks)

In contrast, under double-end feeding, additional disrupted journeys are negligible in most cases, with the largest increase observed under RCP8.5-1,000-year at the 0.3 m threshold, where 8,455 additional disrupted journeys account for only 0.9% of total disrupted journeys. These results indicate that, when double-end feeding redundancy is available, passenger disruption solely attributable to flood-conditioned power dependency remains minimal across the full range of flood scenarios and substation flood-depth thresholds considered. Under single-end feeding, however, accounting for power dependency is particularly essential when lower substation flood-depth thresholds (< 0.9 m) are applied, as a considerable number of substations are exposed to flooding and can then trigger cascading effects that contribute meaningfully to total journey disruption. The detailed morning-peak and afternoon-peak results on satisfied journeys, disrupted journeys, additional disrupted journeys, and corresponding ratios are provided in Appendix A.

### *4.3 Correlation between performance loss and flood-conditioned power dependency*

To further examine whether the proposed flood-conditioned power dependency index reflects network-wide performance loss, Fig. 9 presents the results for each flood scenario across eight modelling configurations, combining two feeding mechanisms with four substation flood-depth thresholds. The results indicate a clear positive correlation, with higher power dependency corresponding to greater performance loss under floods across modelling configurations. This pattern is particularly evident under single-end feeding, where lower substation flood-depth thresholds are associated with both higher dependency exposure and larger reductions in robustness. By contrast, double-end feeding configurations are generally clustered near the origin, indicating limited dependency-related performance loss when feeding redundancy is available.

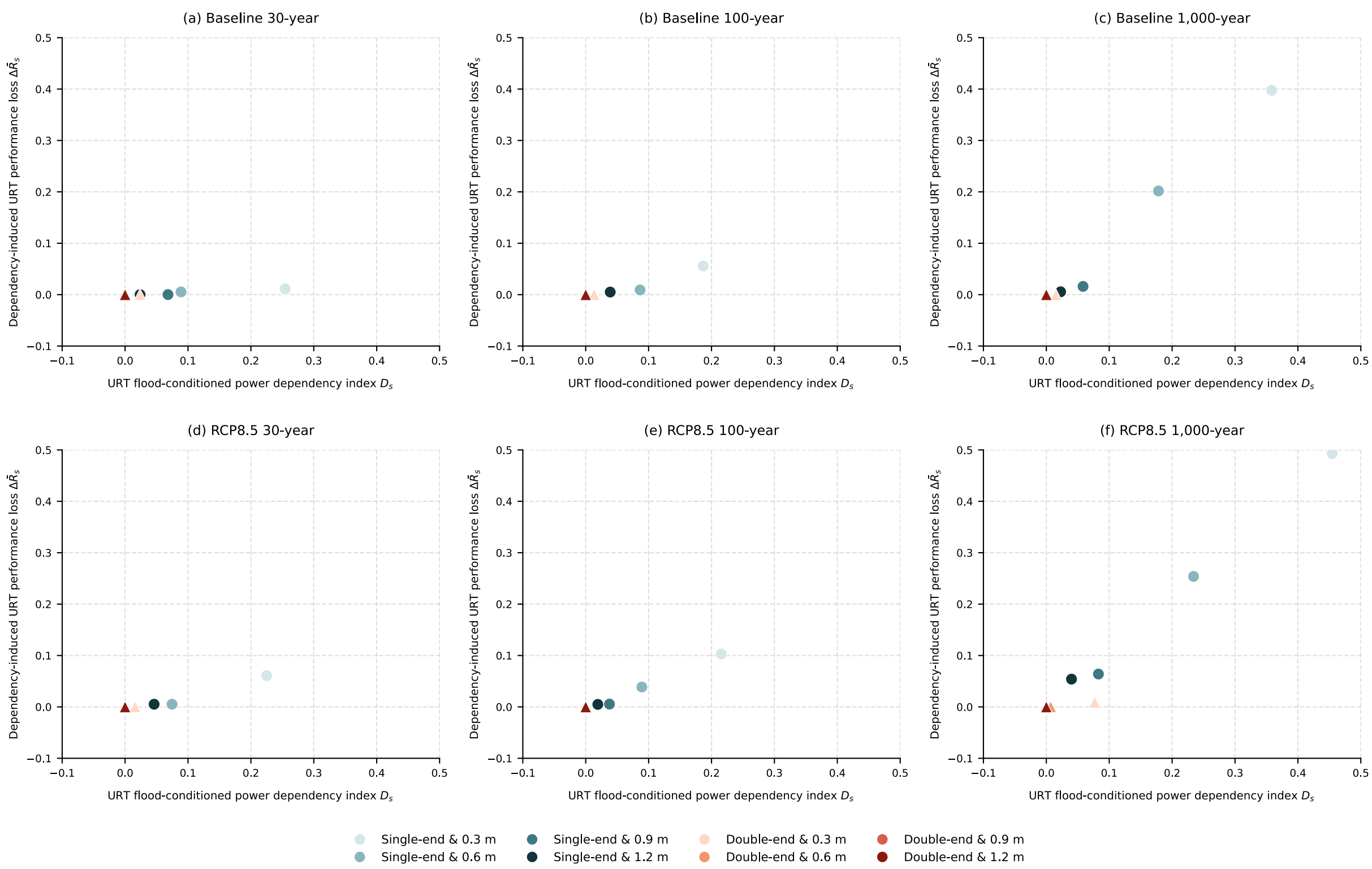

**Fig. 9** Correlation between URT performance loss and power dependency under flood scenarios

Table 3 quantifies these correlations using Spearman's $\rho$. The correlation is consistently strong across all individual flood scenarios, ranging from 0.86 to 1.0, with a pooled $\rho$ of 0.93 across all configurations. The correlation remains robust after excluding zero-value pairs, with $\rho$ ranging from 0.70 to 1.0 and a pooled $\rho$ of 0.83 across 30 configurations. This confirms that the observed relationship is not an artefact of configurations being concentrated at the origin. These results suggest that the flood-conditioned power dependency index provides is an informative, meaningful network-wide indicator of the relative magnitude of URT performance loss attributable to power dependency under flood scenarios.

**Table 3** Spearman's $\rho$ between flood-conditioned power dependency index and loss in URT robustness index (computed per flood scenario and for the pooled dataset)

| **Flood scenario** | **All configurations** | | **Configurations with non-zero results** | |
|---|---|---|---|---|
| | **Spearman's $\rho$** | **No. of configurations** | **Spearman's $\rho$** | **No. of configurations** |
| Baseline: 30-year | 0.86 | 8 | 1.00 | 5 |
| Baseline: 100-year | 1.00 | 8 | 1.00 | 5 |
| Baseline: 1,000-year | 0.98 | 8 | 0.90 | 5 |
| RCP8.5: 30-year | 0.98 | 8 | 0.94 | 6 |
| RCP8.5: 100-year | 1.00 | 8 | 1.00 | 4 |
| RCP8.5: 1,000-year | 0.93 | 8 | 0.70 | 5 |
| Pooled | 0.93 | 48 | 0.83 | 30 |

Note: Each flood scenario includes eight modelling configurations, defined by two feeding mechanisms and four substation flood-depth thresholds. Non-zero configurations refer to cases in which at least one of the two indices is greater than zero.

## 5 Discussion

### *5.1 Traction power redundancy as a critical cascading-failure modelling detail*

The comparison between single-end and double-end feeding mechanisms demonstrates that traction power redundancy fundamentally influences the severity of cascading failures from the URT traction power layer to its operational layer. It should therefore be treated as a key characteristic governing the functional dependency between traction power supply and rail operations, and considered explicitly in cascading-failure modelling. A one-to-one dependency rule, whereby the failure of a substation immediately disables its connected track segment, effectively neglects the redundancy inherent in double-end feeding and can substantially overestimate cascading disruption impacts. More broadly, the findings show that dependency effects are governed not only by the presence of inter-layer links, but also by the failure-propagation rule assigned to those links. Two models with the same assets and dependency connections may therefore produce very different outcomes if they apply different failure-propagation rules.

The London case study corroborates the resilience benefit of double-end-feeding in DC-electrified URT systems. This built-in redundancy, intended to enhance reliability and fault tolerance, limits the propagation of traction power failures, with only minimal power-attributable performance loss even under the most severe RCP8.5-1,000-year scenario. This has practical implications for URT resilience planning: where multiple adverse conditions compete for limited budgets and resources, flood-induced traction power dependency may be assigned a lower priority than risks that produce greater operational impacts. Considering traction power redundancy can therefore help distinguish genuinely critical cascading risks from those that can be largely absorbed by the existing system design, supporting more proportionate prioritisation of resilience strategies.

### *5.2 Sensitivity of cascading disruption to substation flood failure thresholds*

The single-end feeding results reveal a second important modelling detail: cascading impacts are highly sensitive to the flood-depth threshold used to define traction substation failure. When redundancy is absent, this threshold directly affects how many substations are classified as inoperative and how many dependent track segments are consequently disrupted. The sharp decline in both the flood-conditioned power dependency index and dependency-induced performance loss as the threshold increases from 0.3 m to 0.9 m indicates that much of the estimated cascading disruption is associated with relatively shallow flooding. Adopting a single generic failure threshold without accounting for differences in substation design, equipment elevation, local protection, and site conditions could therefore substantially overestimate or underestimate dependency-related risk. Where asset-specific vulnerability data are unavailable, testing a range of thresholds is important not only for representing uncertainty, but also for identifying the level of flood tolerance at which power-related cascading effects are substantially reduced. In the London case, the results suggest that increasing substation flood tolerance towards 0.9 m could substantially reduce cascading disruption under a non-redundant feeding assumption, providing a useful indication of the protection standard that may be required where feeding redundancy is unavailable.

### *5.3 Flood-conditioned power dependency as a proxy for related performance loss*

The strong positive correlation between the flood-conditioned power dependency index and dependency-related performance loss of the studied London URT suggests that the proposed index can serve as a proxy for the operational impacts of flood-induced traction power disruptions. This relationship remains strong across the examined flood scenarios and modelling configurations, including after zero-value cases are excluded, indicating that it is not driven solely by configurations with negligible cascading effects. The index therefore provides more than a count of track segments disrupted by traction power loss; instead, it indicates the extent to which power-side failures contribute to network-wide performance degradation relative to direct inundation. This may enable transport operators to identify scenarios in which traction power dependency is likely to contribute substantially to performance loss without requiring detailed network-disruption and journey-satisfaction simulations. Such an application could be particularly valuable for early-stage resilience assessments or data-scarce settings.

Nevertheless, this relationship is derived from the London case and should not be assumed to apply universally to URT networks in other cities, as the performance consequences of dependency-related failures depend on network topology, passenger-demand distribution, and the operational importance of the affected track segments. The proposed methodology is, however, transferable to other URT networks and can be applied under their specific network topology, inter-layer dependency, travel demand, and hazard conditions, thereby supporting the development of broader evidence on the relationship between hazard-conditioned dependency and dependency-related performance loss.

### *5.4 Interpreting flood-conditioned power dependency under changing flood and climate scenarios*

While future climate scenarios generally increase dependency-related performance loss, the results demonstrate that the relationship between flood severity and flood-conditioned power dependency is more nuanced than a simple monotonic increase. Larger flood extents expose more traction substations and therefore increase the potential for cascading disruption. However, flood-conditioned power dependency is also shaped by the spatial coincidence between directly flooded track segments and those indirectly disrupted through traction power loss. As flood extent expands, increasing overlap between these two types of disruption can reduce the additional impacts attributable solely to traction power loss. Consequently, more severe flood scenarios do not necessarily produce higher dependency-index values, even though their overall resilience impacts are greater.

This finding highlights that flood-conditioned power dependency should be interpreted as a scenario-conditioned property rather than a fixed characteristic of the network. The magnitude of dependency-related impacts depends jointly on network topology, the configuration of the traction power supply system, and the spatial distribution of hazard exposure. Therefore, evaluating infrastructure dependency under climate change requires explicit consideration of how future hazard patterns reshape both direct and indirect disruption pathways, rather than assuming that increasing hazard intensity leads proportionally to greater cascading impacts.

### *5.5 Role of network redundancy and connectivity in absorbing disruptions*

From a network perspective, the results also highlight the role of redundancy and connectivity in determining how local infrastructure failures translate into system-level performance loss. Redundancy operates at two levels in the model. Within the traction power layer, double-end feeding provides an alternative source of power when one supplying substation fails, thereby preventing many local power failures from propagating to the operational layer. Within the URT operational layer, network connectivity provides alternative travel paths when individual track segments become unavailable, allowing some passenger journeys to be rerouted and remain satisfied despite the disruption. The effect of a failed component therefore depends not only on whether it becomes inoperative, but also on whether the surrounding network provides alternative connections capable of maintaining system functionality.

This network redundancy can effectively absorb or redistribute the consequences of local disruptions. Rather than concentrating the effect of a failed track segment on all journeys that originally used it, alternative paths can distribute affected journeys across other parts of the network, thereby reducing the resulting loss of satisfied travel demand. Consequently, similar numbers of inoperative track segments may produce different performance losses depending on their locations and on the availability of alternative routes around them. This also helps explain why component-level disruption alone is insufficient for assessing resilience: the system-level consequence of a failure is shaped by the broader network structure through which its effects can either be contained or accommodated.

## 6 Conclusions

This study developed a novel network model-based methodology to quantify flood-conditioned power dependency of URT and assess its effects on operational performance across varying flood scenarios. The framework integrates a double-layer URT network model, spatial flood exposure assessment, service disruption simulation, a flood-conditioned power dependency index, and a journey-based resilience metric. By distinguishing track failures caused directly by flooding from those attributable solely to traction power loss due to substation flooding, the framework interprets system dependency as a scenario-conditioned property rather than a fixed structural characteristic. It also enables the effects of traction power redundancy and uncertainty in substation flood vulnerability to be examined systematically. The proposed framework could also be adapted to other infrastructure systems, hazards, and interdependency contexts.

The methodology is applied to the DC-electrified lines of the London rail transit network under a range of surface water flood scenarios across current and RCP8.5 climate conditions, two traction power feeding mechanisms, and four substation flood-failure thresholds. The results show that traction power redundancy substantially influences cascading impacts and dependency-related performance loss. Under double-end feeding, the additional performance loss attributable to substation flooding remains limited across the tested scenarios, whereas single-end feeding produced substantially greater impacts and is more sensitive to assumed substation flood-failure thresholds. The fundings also indicate that more severe flooding does not necessarily result in a higher dependency index, because greater spatial overlap between directly flooded tracks and tracks indirectly affected by traction power failures can reduce the additional disruption attributable solely to power loss. Nevertheless, future climate scenarios generally increased overall disruption and dependency-related performance loss. The strong correlation between the proposed dependency index and dependency-related performance loss supports its use as a resilience proxy for strategic planning.

Several limitations provide directions for future research. First, while the proposed methodology captures resilience properties of robustness and redundancy during disruption, it does not explicitly model recovery. Future work could stress-test localised flood-event-based disruption scenarios and model detailed recovery by incorporating repair durations, resource constraints, resource scheduling priority, and coordinated recovery of rail and power assets. Second, because traction substation locations and supply boundaries are security-sensitive information, authoritative data are not publicly available. This study therefore used proxy substation locations to demonstrate the proposed framework. Future work could apply the framework in collaboration with URT operators, enabling the assessment to be performed using confidential operational data while allowing the results to be validated without disclosing sensitive asset information. Third, although the findings derived from the London URT case provide useful evidence on the mechanisms through which flood-conditioned power dependency influences URT resilience, their quantitative implications should not be generalised directly to other URT networks. Future research should apply the framework across URT networks with different structural, operational, and climatic characteristics to assess the transferability of the findings and identify more general patterns. Finally, this study considers only the dependency of track operation on traction power and does not capture the dependence of stations on auxiliary power supply. Station flooding and the resulting loss of operational performance are therefore outside the scope of this study. Future research should incorporate auxiliary power–station dependencies to provide a more comprehensive assessment framework of URT–power dependencies.

### CRediT authorship contribution statement

**Wei Bi:** Conceptualization, Data curation, Methodology, Formal analysis, Writing – original draft, Writing – review & editing, Funding acquisition. **Jürgen Hackl:** Writing – review & editing. **Bryan T. Adey:** Writing – review & editing, Funding acquisition.

### Declaration of competing interest

The authors declare that they have no known competing financial interests or personal relationships that could have appeared to influence the work reported in this paper.

### Data availability

Data will be made available on request.

### Acknowledgement

This study was supported by an ETH Zurich Postdoctoral Fellowship.

## Appendix A Statistics of peak-hour journey impacts under substation flood cascading effects

**Table A. 1** Peak-hour journey impacts under substation-related cascading effects (single-end feeding)

| Substation flood depth threshold | Flood scenario | AM-peak satisfied journeys | AM-peak satisfied journey ratio $R^{\mathrm{AM}}$ | AM-peak disrupted journeys | Disrupted AM-peak journeys due to substation cascading effects | $\Delta R^{\mathrm{AM}}$ | PM-peak satisfied journeys | PM-peak satisfied journey ratio $R^{\mathrm{PM}}$ | PM-peak disrupted journeys | Disrupted PM-peak journeys due to substation cascading effects | $\Delta R^{\mathrm{PM}}$ | $\overline{R}$ | $\overline{\Delta R}$ |
|---|---|---|---|---|---|---|---|---|---|---|---|---|---|
| 0.3 m | Baseline: 30-year | 632,579 | 73.3% | 230,871 | 10,437 | 1.2% | 758,310 | 75.5% | 245,732 | 10,696 | 1.1% | 75.5% | 1.1% |
| | Baseline: 100-year | 491,009 | 56.9% | 372,442 | 48,390 | 5.6% | 601,801 | 59.9% | 402,241 | 55,399 | 5.5% | 59.9% | 5.5% |
| | Baseline: 1,000-year | 126,434 | 14.6% | 737,017 | 341,180 | 39.5% | 174,734 | 17.4% | 829,308 | 401,508 | 40.0% | 17.4% | 40.0% |
| | RCP8.5: 30-year | 557,963 | 64.6% | 305,487 | 53,228 | 6.2% | 674,452 | 67.2% | 329,590 | 60,244 | 6.0% | 67.2% | 6.0% |
| | RCP8.5: 100-year | 416,959 | 48.3% | 446,491 | 87,263 | 10.1% | 510,520 | 50.8% | 493,522 | 105,549 | 10.5% | 50.8% | 10.5% |
| | RCP8.5: 1,000-year | 31,085 | 3.6% | 832,365 | 415,832 | 48.2% | 47,262 | 4.7% | 956,780 | 506,300 | 50.4% | 4.7% | 50.4% |
| 0.6 m | Baseline: 30-year | 632,579 | 73.9% | 225,216 | 4,781 | 0.6% | 763,792 | 76.1% | 240,251 | 5,215 | 0.5% | 76.1% | 0.5% |
| | Baseline: 100-year | 491,009 | 61.5% | 332,263 | 8,211 | 1.0% | 648,383 | 64.6% | 355,659 | 8,817 | 0.9% | 64.6% | 0.9% |
| | Baseline: 1,000-year | 126,434 | 33.9% | 570,906 | 175,069 | 20.3% | 374,568 | 37.3% | 629,474 | 201,674 | 20.1% | 37.3% | 20.1% |
| | RCP8.5: 30-year | 557,963 | 70.3% | 256,863 | 4,603 | 0.5% | 729,621 | 72.7% | 274,421 | 5,075 | 0.5% | 72.7% | 0.5% |
| | RCP8.5: 100-year | 416,959 | 54.6% | 391,647 | 32,419 | 3.8% | 576,075 | 57.4% | 427,967 | 39,994 | 4.0% | 57.4% | 4.0% |
| | RCP8.5: 1,000-year | 31,085 | 26.8% | 632,473 | 215,940 | 25.0% | 294,917 | 29.4% | 709,125 | 258,645 | 25.8% | 29.4% | 25.8% |
| 0.9 m | Baseline: 30-year | 642,880 | 74.5% | 220,570 | 136 | 0% | 768,948 | 76.6% | 235,094 | 59 | 0% | 76.6% | 0% |
| | Baseline: 100-year | 534,925 | 62.0% | 328,526 | 4,474 | 0.5% | 652,247 | 65.0% | 351,795 | 4,953 | 0.5% | 65.0% | 0.5% |
| | Baseline: 1,000-year | 453,882 | 52.6% | 409,569 | 13,732 | 1.6% | 559,972 | 55.8% | 444,070 | 16,270 | 1.6% | 55.8% | 1.6% |
| | RCP8.5: 30-year | 606,588 | 70.3% | 256,863 | 4,603 | 0.5% | 729,621 | 72.7% | 274,421 | 5,075 | 0.5% | 72.7% | 0.5% |
| | RCP8.5: 100-year | 499,390 | 57.8% | 364,060 | 4,832 | 0.6% | 610,823 | 60.8% | 393,220 | 5,247 | 0.5% | 60.8% | 0.5% |
| | RCP8.5: 1,000-year | 392,582 | 45.5% | 470,868 | 54,335 | 6.3% | 488,303 | 48.6% | 515,739 | 65,259 | 6.5% | 48.6% | 6.5% |
| 1.2 m | Baseline: 30-year | 642,970 | 74.5% | 220,480 | 46 | 0% | 768,983 | 76.6% | 235,060 | 24 | 0% | 76.6% | 0% |
| | Baseline: 100-year | 534,925 | 62.0% | 328,526 | 4,474 | 0.5% | 652,247 | 65.0% | 351,795 | 4,953 | 0.5% | 65.0% | 0.5% |
| | Baseline: 1,000-year | 462,829 | 53.6% | 400,621 | 4,784 | 0.6% | 570,671 | 56.8% | 433,371 | 5,571 | 0.6% | 56.8% | 0.6% |
| | RCP8.5: 30-year | 606,588 | 70.3% | 256,863 | 4,603 | 0.5% | 729,621 | 72.7% | 274,421 | 5,075 | 0.5% | 72.7% | 0.5% |
| | RCP8.5: 100-year | 499,915 | 57.9% | 363,535 | 4,307 | 0.5% | 611,298 | 60.9% | 392,744 | 4,771 | 0.5% | 60.9% | 0.5% |
| | RCP8.5: 1,000-year | 400,922 | 46.4% | 462,528 | 45,995 | 5.3% | 498,456 | 49.6% | 505,586 | 55,106 | 5.5% | 49.6% | 5.5% |

**Table A. 2** Peak-hour journey impacts under substation-related cascading effects (double-end feeding)

| Substation flood depth threshold | Flood scenario | AM-peak satisfied journeys | AM-peak satisfied journey ratio $R^{\mathrm{AM}}$ | AM-peak disrupted journeys | Disrupted AM-peak journeys due to substation cascading effects | $\Delta R^{\mathrm{AM}}$ | PM-peak satisfied journeys | PM-peak satisfied journey ratio $R^{\mathrm{PM}}$ | PM-peak disrupted journeys | Disrupted PM-peak journeys due to substation cascading effects | $\Delta R^{\mathrm{PM}}$ | $\bar{R}$ | $\overline{\Delta R}$ |
|---|---|---|---|---|---|---|---|---|---|---|---|---|---|
| 0.3 m | Baseline: 30-year | 642,682 | 74.4% | 220,769 | 334 | 0.04% | 768,676 | 76.6% | 235,366 | 330 | 0.03% | 75.5% | 0.04% |
| | Baseline: 100-year | 539,281 | 62.5% | 324,169 | 117 | 0.01% | 657,078 | 65.4% | 346,964 | 122 | 0.01% | 63.9% | 0.01% |
| | Baseline: 1,000-year | 467,614 | 54.2% | 395,837 | 0 | 0% | 576,242 | 57.4% | 427,800 | 0 | 0% | 55.8% | 0% |
| | RCP8.5: 30-year | 610,857 | 70.7% | 252,593 | 334 | 0.04% | 734,366 | 73.1% | 269,677 | 330 | 0.03% | 71.9% | 0.04% |
| | RCP8.5: 100-year | 504,223 | 58.4% | 359,228 | 0 | 0% | 616,069 | 61.4% | 387,973 | 0 | 0% | 59.9% | 0% |
| | RCP8.5: 1,000-year | 439,300 | 50.9% | 424,151 | 7,618 | 0.88% | 544,270 | 54.2% | 459,772 | 9,292 | 0.93% | 52.5% | 0.90% |
| 0.6 m | Baseline: 30-year | 643,016 | 74.5% | 220,435 | 0 | 0% | 769,006 | 76.6% | 235,036 | 0 | 0% | 75.5% | 0% |
| | Baseline: 100-year | 539,398 | 62.5% | 324,052 | 0 | 0% | 657,200 | 65.5% | 346,842 | 0 | 0% | 64.0% | 0% |
| | Baseline: 1,000-year | 467,614 | 54.2% | 395,837 | 0 | 0% | 576,242 | 57.4% | 427,800 | 0 | 0% | 55.8% | 0% |
| | RCP8.5: 30-year | 611,191 | 70.8% | 252,259 | 0 | 0% | 734,696 | 73.2% | 269,346 | 0 | 0% | 72.0% | 0% |
| | RCP8.5: 100-year | 504,223 | 58.4% | 359,228 | 0 | 0% | 616,069 | 61.4% | 387,973 | 0 | 0% | 59.9% | 0% |
| | RCP8.5: 1,000-year | 446,649 | 51.7% | 416,802 | 269 | 0.03% | 553,119 | 55.1% | 450,923 | 443 | 0.04% | 53.4% | 0.04% |
| 0.9 m | Baseline: 30-year | 643,016 | 74.5% | 220,435 | 0 | 0% | 769,006 | 76.6% | 235,036 | 0 | 0% | 75.5% | 0% |
| | Baseline: 100-year | 539,398 | 62.5% | 324,052 | 0 | 0% | 657,200 | 65.5% | 346,842 | 0 | 0% | 64.0% | 0% |
| | Baseline: 1,000-year | 467,614 | 54.2% | 395,837 | 0 | 0% | 576,242 | 57.4% | 427,800 | 0 | 0% | 55.8% | 0% |
| | RCP8.5: 30-year | 611,191 | 70.8% | 252,259 | 0 | 0% | 734,696 | 73.2% | 269,346 | 0 | 0% | 72.0% | 0% |
| | RCP8.5: 100-year | 504,223 | 58.4% | 359,228 | 0 | 0% | 616,069 | 61.4% | 387,973 | 0 | 0% | 59.9% | 0% |
| | RCP8.5: 1,000-year | 446,917 | 51.8% | 416,533 | 0 | 0% | 553,562 | 55.1% | 450,480 | 0 | 0% | 53.4% | 0% |
| 1.2 m | Baseline: 30-year | 643,016 | 74.5% | 220,435 | 0 | 0% | 769,006 | 76.6% | 235,036 | 0 | 0% | 75.5% | 0% |
| | Baseline: 100-year | 539,398 | 62.5% | 324,052 | 0 | 0% | 657,200 | 65.5% | 346,842 | 0 | 0% | 64.0% | 0% |
| | Baseline: 1,000-year | 467,614 | 54.2% | 395,837 | 0 | 0% | 576,242 | 57.4% | 427,800 | 0 | 0% | 55.8% | 0% |
| | RCP8.5: 30-year | 611,191 | 70.8% | 252,259 | 0 | 0% | 734,696 | 73.2% | 269,346 | 0 | 0% | 72.0% | 0% |
| | RCP8.5: 100-year | 504,223 | 58.4% | 359,228 | 0 | 0% | 616,069 | 61.4% | 387,973 | 0 | 0% | 59.9% | 0% |
| | RCP8.5: 1,000-year | 446,917 | 51.8% | 416,533 | 0 | 0% | 553,562 | 55.1% | 450,480 | 0 | 0% | 53.4% | 0% |